\documentclass[twocolumn,secnumarabic,amssymb, nobibnotes, aps, prd]{revtex4-1}

\usepackage{graphicx}
\usepackage{subfigure}
\begin{document}

\title{Complete electrodynamics of a BCS superconductor with $\mu$eV energy scales: microwave spectroscopy on titanium at mK temperatures}%

\author{Markus Thiemann, Martin Dressel, Marc Scheffler}%
\affiliation{1. Physikalisches Institut, Universit\"at Stuttgart, Pfaffenwaldring 57, D-70569 Stuttgart, Germany}
\date{\today}%
\begin{abstract}

We performed resonant microwave measurements on superconducting titanium (Ti) down to temperatures of 40~mK, well below its critical temperature $T_\mathrm{c} \approx 0.5$~K. Our wide frequency range  3.3-40~GHz contains the zero-temperature energy gap $2\Delta_0$ and allows us to probe the full electrodynamics of the superconducting state, including excitations across the gap and the low-frequency responses of superfluid condensate and thermal quasiparticles. The observed behavior follows the predictions of the BCS-based Mattis-Bardeen formalism, which implies that superconducting Ti is in the dirty limit, in agreement with our determination of the scattering rate.
We directly determine the temperature dependence of the energy gap, which is in accordance with BCS predictions, and $2\Delta_0/k_\mathrm{B}T_\mathrm{c}\approx3.5$ with $\Delta_0 \approx$ 75~$\mu$eV.
We also evaluate the penetration depth, and we characterize the behavior of superconducting Ti in external magnetic field.
\end{abstract}
\maketitle

\section{\label{sec:intro}Introduction}
Optical spectroscopy is a versatile tool to investigate the fundamental electronic characteristics of superconductors \cite{Basov2005, Pracht2013}: single-particle excitations indicate the superconducting energy gap; the superfluid condensate and the penetration depth are probed via the out-of phase response; and the quasiparticle dynamics are sensed via sub-gap absorption. These experimental virtues have lead to such groundbreaking results as the first observations of the superconducting energy gap $2\Delta$ via far-infrared spectroscopy and of the condensate reactance via microwave studies of Pb and Sn \cite{TinkhamPhysRev1957,TinkhamPhysRev1960,Dressel_AdvCondMatPhys2013} around the time that BCS theory was developed \cite{Bardeen_PhysRev1952}. Microwave measurements also gave first robust evidence in the 1990's for the linear temperature dependence of the penetration depth in cuprate superconductors, which suggests d-wave symmetry of the superconducting order parameter \cite{HardyPRL1993,Gao_ApplPhysLett,Bonn_PhysRevB1993}. Consequently, numerous superconducting materials have been studied with electrodynamic experiments in the infrared, THz, or microwave spectral range \cite{Basov2005, Maeda2005, Charnukha2015}.
Most of these experiments were performed at temperatures of liquid $^4$He, whereas only very few optical studies addressed temperatures below 1~K \cite{Basov2003, Crane2007, HosseinKhah2010, Liu2011, Steinberg2012, Driessen2012}. 
Experimental challenges for a long time precluded electrodynamic studies of superconductors at ultralow temperatures \cite{Scheffler2017}, and thus all superconductors with critical temperature $T_\mathrm{c}$ well below 1~K could not be probed by optics, with microwave spectroscopy being particularly relevant (thermal energy $k_BT$ for 1~K corresponds to 86~$\mu$eV photon energy $\hbar\omega$ or 21~GHz).
Considering the wide range of unconventional low-$T_\mathrm{c}$ superconductors that are presently studied at mK temperatures with other techniques, spanning heavy-fermion superconductors \cite{Pfleiderer2009}, materials near a superconductor-insulator transition \cite{Gantmakher2010, Pracht2016}, ultra-low density superconductors \cite{Behnia2017}, or the LaAlO$_3$/SrTiO$_3$ interface \cite{Caviglia2008}, the lack of optical data is quite unfortunate.

Recent experimental advances now allow microwave spectroscopy experiments in $^3$He/$^4$He dilution refrigerators \cite{Ormeno2002, Ormeno2006, Truncik_NatCom2013, Scheffler2015, Wiemann2015, Li_NewJPhys2016, Scheffler2017}, and thus the microwave response at mK temperatures of such superconducting materials with rather low $T_\mathrm{c}$ comes into focus. 
These experiments operate in a previously unexplored regime, considering that the accessible GHz spectral range includes frequencies both smaller and larger than 2$\Delta$ of mK superconductors \cite{Scheffler2017} and that $T_\mathrm{c}$ or $2\Delta$ can be much smaller than other energy scales, e.g.\ the scattering rate.
While this newly accessible experimental regime prompts studies on numerous exotic superconducting states, at the same time it calls for investigations of superconductors with $T_\mathrm{c}$ well below 1~K that are considered conventional superconductors and thus allow investigations of BCS-like behavior in previously inaccessible parameter ranges and that at the same time can act as references for similar experiments on unconventional low-$T_\mathrm{c}$ superconductors.
This is our motivation to choose the elemental superconductor Ti with $T_\mathrm{c}$ around 0.5~K \cite{Daunt1949, Smith_PhysRev1952, Steele_PhysRev1953, Peruzzi_NucPhysB1999, Peruzzi_Metr2000} for this investigation of the complete electrodynamics of a mK superconductor.
The role of sample purity for superconductivity in Ti is evident from early experiments \cite{Meissner_ZfP1930, Shoenberg_Proc1940} as well as more detailed recent work \cite{Peruzzi_NucPhysB1999, Peruzzi_Metr2000}. Furthermore, de Haas-van Alphen measurements on Ti indicate different Fermi sheets with effective masses ranging from $m^*=1m_e$ to $m^*=3m_e$ \cite{Kamm_LT131974,Welch_PRB1974}, making titanium a candidate for multiband superconductivity, like recently observed for another elemental superconductor, Pb \cite{Ruby2015}.
Superconductivity in Ti is also exploited in various mK devices \cite{Heersche2007, Sacepe2011, Vissers2013}.

We employ a microwave multimode resonator to obtain the optical conductivity of superconducting Ti for frequencies and temperatures ranging from $3-40$~GHz and $40-600$~mK, respectively, smoothly crossing from $\hbar\omega\ll k_BT$ to $\hbar\omega\gg k_BT$. This allows us to measure across the superconducting energy gap and observe the temperature dependence of the gap as a unique feature in our optical conductivity spectra. Furthermore, we evaluate the full electrodynamic response of Ti within the BCS framework, and we investigate the superconducting state of Ti in an external magnetic field.

\begin{figure}
\includegraphics[width=0.8\linewidth]{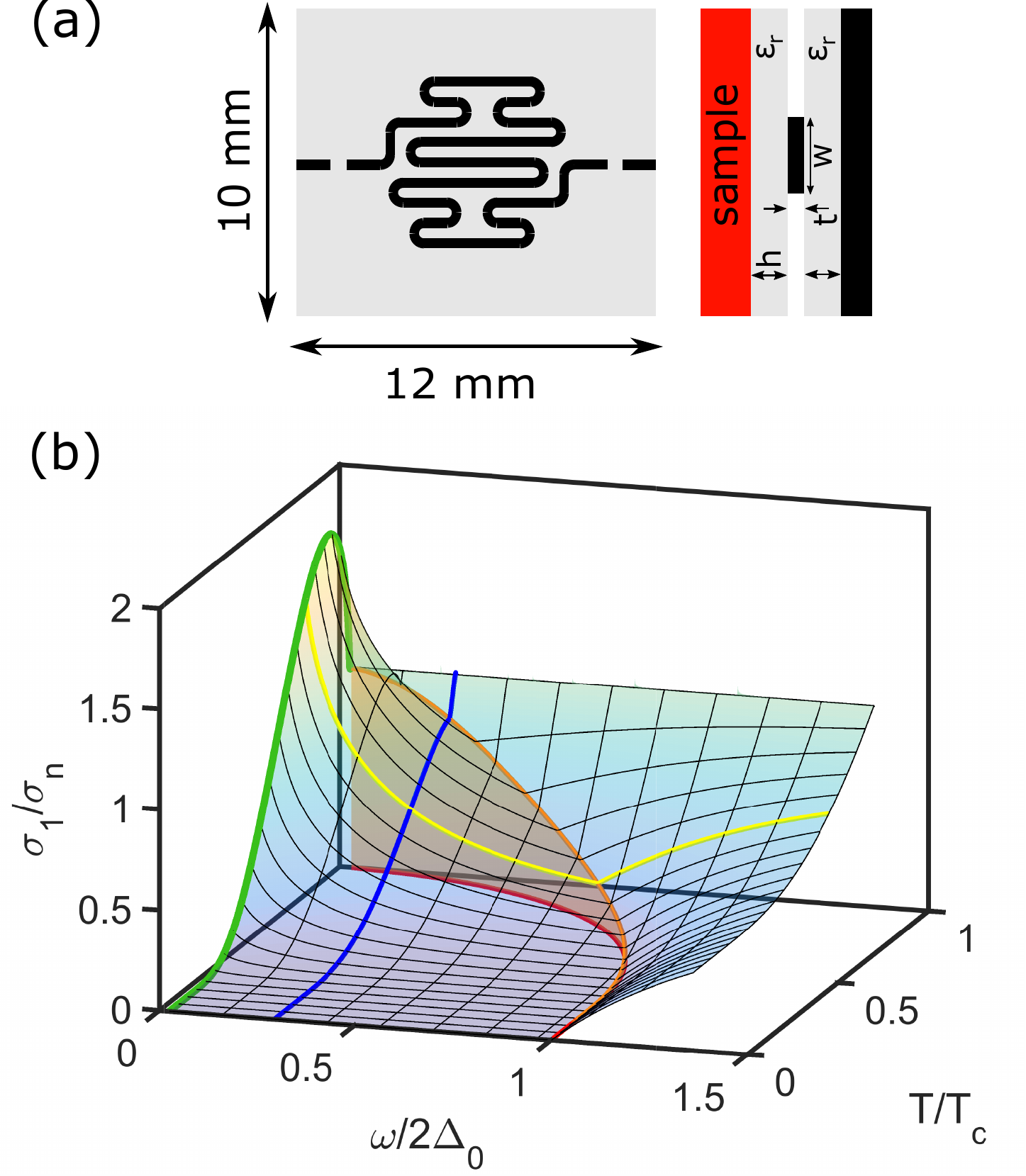}
\caption{(a) Schematic of a stripline resonator. (left) Top view on the center conductor. (right) Cross section with relevant parameters:
dielectric constant $\epsilon_r$ and thickness $h$ of the dielectric, center conductor thickness $t$ and width $w$. (b) Real part of the optical conductivity $\sigma_1/\sigma_n$ as a function of excitation frequency and temperature, calculated based on the Mattis-Bardeen formalism \cite{Mattis_PhysREv1958}. The blue and yellow lines indicate $\sigma_1(T)$ and $\sigma_1(\omega)$ at constant frequency and temperature, respectively. The orange line marks the abrupt change in $\sigma_1$, at the combination of temperature and frequency where the excitation frequency matches the energy gap $2\Delta$. By projecting this line to the frequency - temperature plane one can track the magnitude and temperature dependence of the energy gap, which is shown as red line. The green line shows the coherence peak at low frequencies.}
\label{pic:stripline}
\end{figure}

\section{\label{sec:exp}Experiment}
To combine microwave spectroscopy with mK temperatures, we employ superconducting stripline resonators \cite{DiIorio_PRB1988, Scheffler2012, Scheffler2013, Hafner_RevSciInstr2014}. A stripline is formed by a planar center conductor, sandwiched between two dielectrics followed by two ground planes, as shown in the schematic drawing of Fig.~\ref{pic:stripline}(a). The center conductor has a meandered shape to increase its length beneath the sample. This allows us to achieve fundamental frequencies of about 1.5~GHz. The gaps in the center conductor define a one-dimensional resonant structure, with harmonics spaced equally in frequency. By measuring several of the harmonics, we gain information about the frequency dependence. In the used stripline geometry the sample acts as ground plane, and therefore a change in the microwave properties (i.e. the optical conductivity at GHz frequencies) of the sample acts as a perturbation on the resonator. This results in a shift of the resonance frequency $f_0$ and a change in the resonant bandwidth $f_B$ compared to the unperturbed ideal resonator. The measured quantities $f_B$ and $f_0$ can be related to the surface impedance $Z_s=R_s-iX_s$ of the sample via cavity perturbation theory~\cite{Klein_cavper1993}:
\begin{equation}
R_s-i\Delta X_s=G\left(\frac{f_B^\mathrm{sample}}{2}-i\Delta f_0 \right)
\label{eq:perturbation} 
\end{equation}
Here $\Delta f_0$ is the change in the resonance frequency that is caused by the sample compared to an unperturbed resonator. $G$ is the resonator constant, which depends on the resonator geometry and the interaction of the electromagnetic fields with the sample. 
To determine $\Delta f_0(T)$ from the experimentally measured frequencies $f_0(T)$, we have to know the absolute value of $\Delta f_0$ for one reference temperature, and to this end we assume that $R_s$ and $X_s$ match at temperatures above $T_\mathrm{c}$, in the metallic state, and we introduce the appropriate additive constant to the $X_s$ data.
This procedure is valid for frequencies below the scattering rate of the sample, i.e.\ in the Hagen-Rubens regime \cite{Dressel_Buch}. Assuming local electrodynamics, where the mean free path of the electrons is shorter than the skin depth, we can then calculate the optical conductivity $\sigma=\sigma_1+i\sigma_2$ via \cite{Dressel_Buch}:
\begin{equation}
\sigma=\frac{i\omega\mu_0}{Z_s^2}
\label{eq:sigma}
\end{equation}

The dimensions of the stripline (see schematic cross section in Fig.~\ref{pic:stripline}(a)) are as follows to match the characteristic impedance of $50$~$\Omega$ of the external microwave circuitry \cite{Wheeler_IEEE1978}: 
thickness $h=127$~$\mu$m and dielectric constant $\epsilon_r\approx 10$ of the dielectric, width $w=50$~$\mu$m and thickness $t=1$~$\mu$m of the center conductor. The gaps in the inner conductor, which define the length of the resonator, were $100$~$\mu$m wide to provide appropriate coupling.

To be as sensitive as possible to the sample of interest, the internal losses of the resonator have to be minimized. Therefore we use sapphire as a dielectric due to its low microwave losses \cite{Krupka_MesSciTech1999}. The conductive parts of the resonator, colored black in Fig~\ref{pic:stripline}(a), are made of superconducting Pb with a $T_\mathrm{c}\approx 7.2$~K. The center conductor is formed by thermal evaporation using a shadow mask.  With pure Pb resonators, where the sample is Pb as well, quality factors exceeding $10^5$ can be achieved \cite{Ebensperger2016}.
The resonator is mounted in a brass box, which is directly connected to the coldfinger of a commercial dilution refrigerator.

The Ti sample (dimensions: $9.5 \times 9.5 \times 1$ mm$^3$) was cut from a Ti plate with purity of 99.999\% \cite{TiSample}.
The inset of Fig.~\ref{pic:sup_density} shows the temperature dependence of the DC-resistivity $\rho_\mathrm{DC}$ of a separate sample with dimensions of $10 \times 1 \times 1$ mm$^3$ cut from the same Ti plate, measured in four-point geometry in a $^4$He cryostat. 
The comparably low residual resistance ratio (RRR) value of 23.9 and the flattening of $\rho_\mathrm{DC}$ at around 30~K indicate substantial defect scattering present in the sample. Using the plasma frequency given in Ref.\ \cite{Ordal_ApplOpt1985} $\omega_p=20300$~cm$^{-1}$ and $\Gamma_\rho=\epsilon_0\omega_p^2\rho_\mathrm{DC}$, we can estimate the scattering rate to $\Gamma_\rho/2\pi=490$~GHz. This value is well above our measurement frequencies and thus justifies the assumption of our sample being in the Hagen-Rubens regime.

We performed mK microwave measurements on the same sample twice, and we could easily determine $T_\mathrm{c}$ from a sharp drop in the resonant bandwidth $f_B(T)$ at the lowest frequency. In the first measurement, we observed $T_\mathrm{c}\approx 0.47$~K. Then the sample was polished, and in the second measurement we found $T_\mathrm{c} \approx 0.50$~K.

\section{Data analysis}
Superconducting Pb resonators bear loss mechanisms that originate from the polycrystalline structure of the evaporated Pb, defects and oxides on the conducting surfaces, and coupling losses \cite{Attanasio1991, Pierce1973}. 
In our temperature range $T<1$~K, these effects may depend on frequency, but are usually temperature independent. Due to Eq.~(\ref{eq:perturbation}) the residual losses can be expressed in terms of a bandwidth $f_B^\mathrm{res}$ adding to the bandwidth $f_B^\mathrm{sample}$ caused by the intrinsic losses of the sample. The measured bandwidth can then be expressed by $f_B(T)=f_B^\mathrm{sample}(T)+f_B^\mathrm{res}$ allowing us to determine $f_B^\mathrm{res}=f_B(T_0)$, where $T_0$ denotes the lowest measured temperature. At low temperatures, the losses of a superconductor become very small and $f_B^\mathrm{sample}(T)\ll f_B^\mathrm{res}$. $f_B^\mathrm{sample}(T)$ can then be calculated via $f_B^\mathrm{sample}(T)=f_B(T)-f_B(T_0)=f_B(T)-f_B^\mathrm{res}$. This procedure is certainly valid for fully gapped superconductors with a $T_\mathrm{c}$ which is a few times higher than the lowest measurable temperature and frequencies below the energy gap $2\Delta_0 = 2 \Delta(T=0)$. The surface impedance is then determined from $f_B^\mathrm{sample}$ via Eq.~\ref{eq:perturbation}.

\begin{figure}
	\includegraphics[width=1\linewidth]{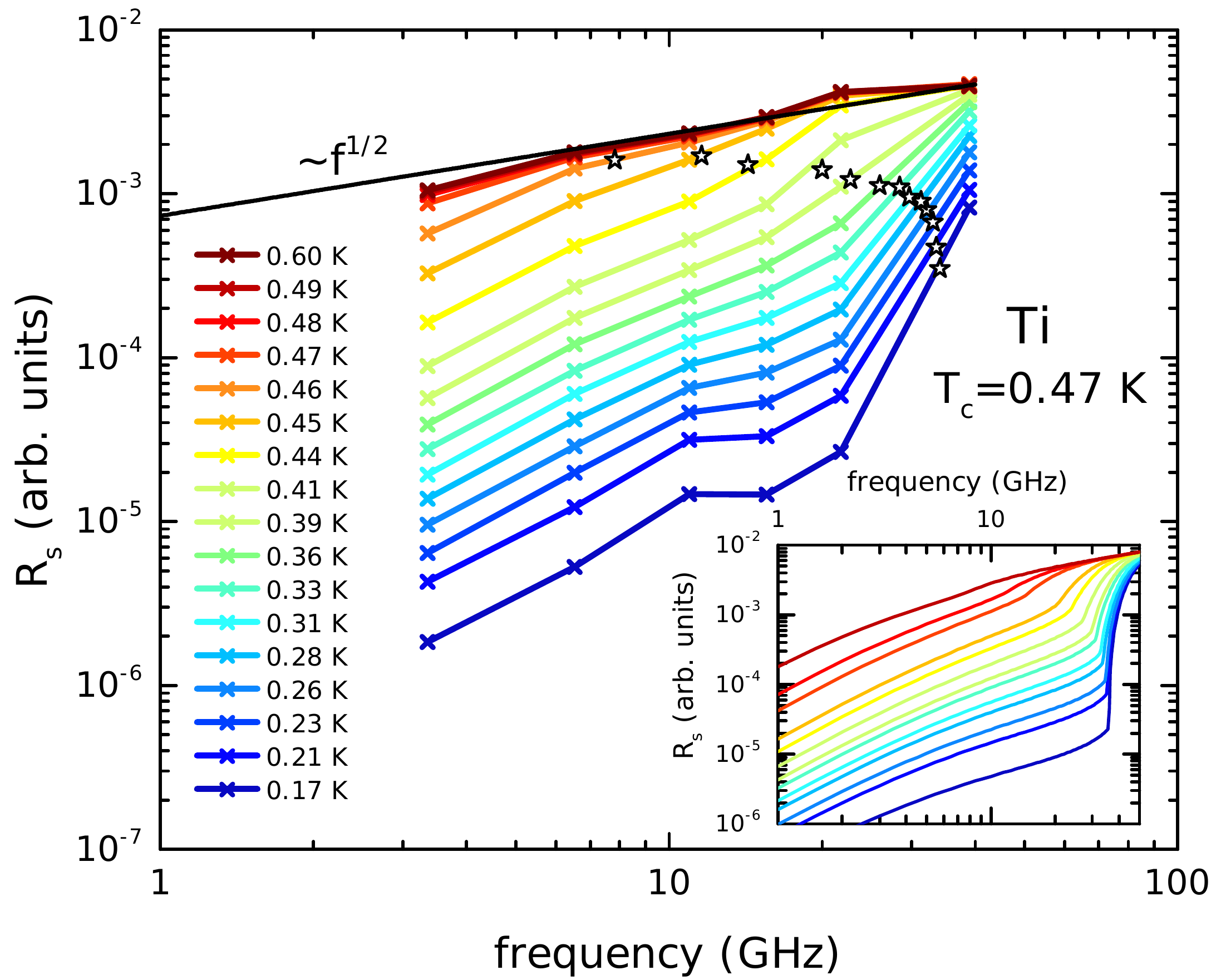}
	\caption{Frequency dependence of the surface resistance of Ti for temperatures across $T_\mathrm{c}$. The black line is a square root fit to the lowest five frequencies for $T =$ 0.60~K, well above $T_\mathrm{c}$. 
In the superconducting state a change in the frequency dependence is visible when crossing the energy gap $2 \Delta$. The open stars denote the frequency of the theoretically expected energy gap $2\Delta(T)$ at each temperature. (Inset) Calculated surface resistance using the Mattis-Bardeen formalism with a $T_\mathrm{c}=0.47$~K and $2\Delta_0/k_\mathrm{B}T_\mathrm{c}=3.53$. }
	\label{pic:Rsvsf}
\end{figure}

In Fig.~\ref{pic:Rsvsf} the surface resistance $R_s(\omega)$ for different temperatures in the superconducting as well as in the metallic state are shown. For the first five frequencies, we applied the above correction procedure. 
But for the highest frequency we cannot expect that the condition $f_B^\mathrm{sample}(T)\ll f_B^\mathrm{res}$ is fulfilled, since the frequency is near the expected energy gap $2\Delta_0/h=3.53k_\mathrm{B}T_\mathrm{c} /h=34$~GHz of Ti, and absorption by breaking Cooper pairs is possible even at the lowest temperature. Therefore we use the frequency dependence $R_s\propto \omega^{1/2}$, expected in the Hagen-Rubens-regime above $T_\mathrm{c}$, to correct the bandwidth at this frequency \cite{Reuter_RoySoc1948}.
The black line in Fig.~\ref{pic:Rsvsf} is a square root fit to the normal-state $R_s$ for the first five frequencies (3.35~GHz to 21.79~GHz), lying below the energy gap. We extrapolate the fit, and match the surface resistance at the highest measured frequency to the extrapolated value. 

The inset of Fig.~\ref{pic:Rsvsf} shows the expected frequency dependence of $R_s$ of Ti, calculated within the Mattis-Bardeen formalism \cite{Mattis_PhysREv1958} and assuming an energy gap $2\Delta_0=3.53k_\mathrm{B}T_\mathrm{c} \;\widehat{=}\; 34$~GHz. Clearly, the phenomenology of frequency- and temperature-dependent $R_s$ observed in our data match these theoretical expectations. In particular, the sharp rise in $R_s(\omega)$ around 40~GHz for lowest temperature and moving to lower frequencies with increasing temperature marks the energy gap $2\Delta/h$.  
In Fig.~\ref{pic:Rsvsf} the frequency that corresponds to $2\Delta(T)$ is indicated by the black edged stars for the different temperatures.

\section{Results and discussion}

\begin{figure*}
	\includegraphics[width=1\linewidth]{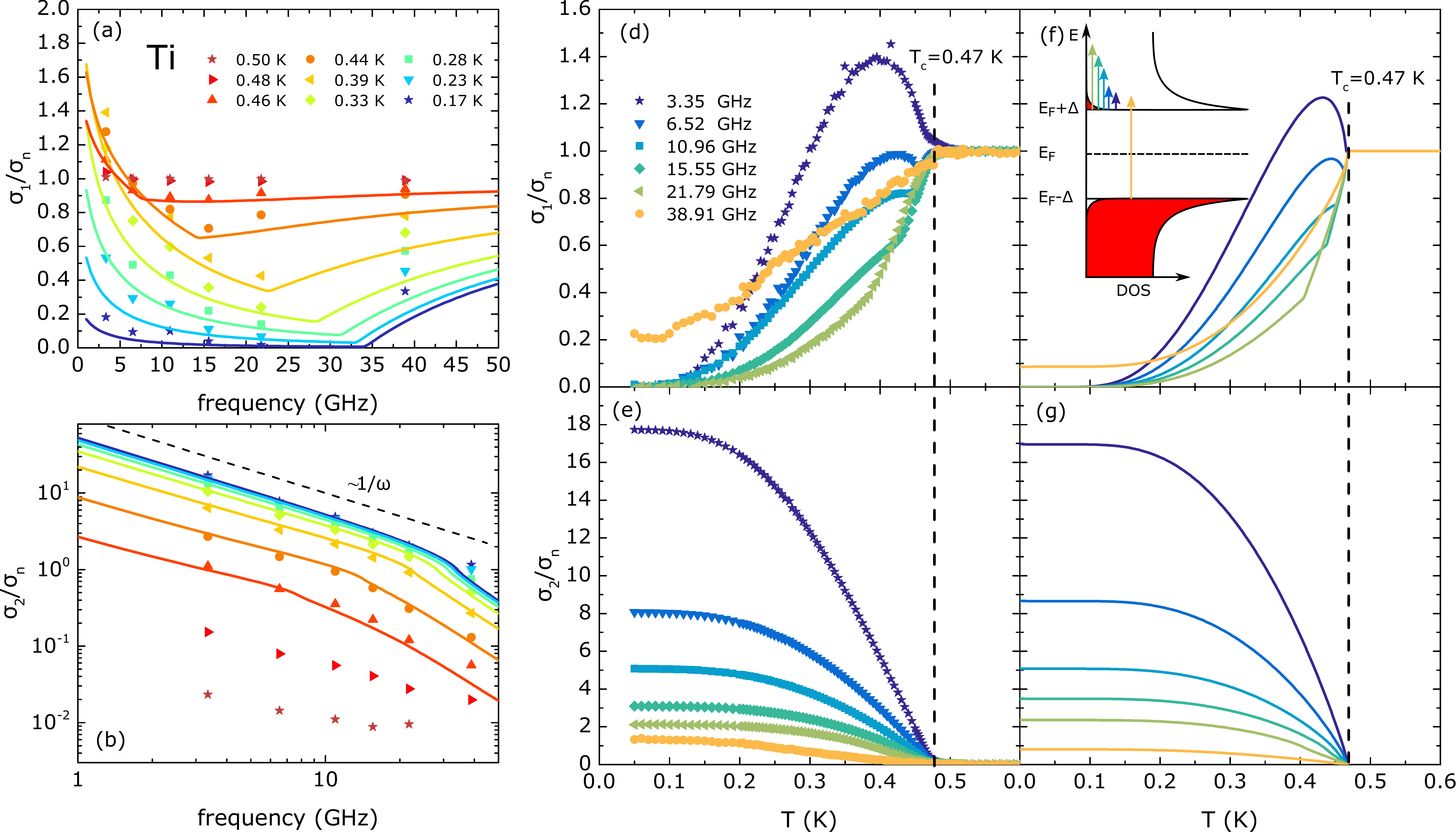}
	\caption{(a),(b) Frequency dependence of $\hat{\sigma}=\sigma_1+\mathrm{i}\sigma_2$ at various temperatures. The solid lines represent the frequency behavior calculated from the Mattis-Bardeen equations using a $T_\mathrm{c}=0.47$~K and $2\Delta_0/k_\mathrm{B}T_\mathrm{c}=3.53$. The energy gap appears in $\sigma_1 (\omega)$ as a kink. (c),(d) Temperature dependence of $\sigma$ at different measured frequencies. (e),(f) Calculated optical conductivity using the Mattis-Bardeen formalism with $T_\mathrm{c}=0.47$~K and $2\Delta_0/k_\mathrm{B}T_\mathrm{c}=3.53$. The inset is a depiction of the density of states (DOS). The colored arrows indicate the possible excitations at the different frequencies.}	
	\label{pic:sigma_main}
\end{figure*}

\subsection{Electrodynamics of the superconducting state}

The interpretation of optical spectra measured on superconductors is usually done with respect to the complex optical conductivity $\sigma=\sigma_1+\mathrm{i}\sigma_2$, where $\sigma_1$ is connected to the absorption rate and $\sigma_2$ to the phase shift of the electromagnetic wave. For dirty superconductors, with scattering rate $\Gamma$ much larger than the superconducting energy gap $2\Delta/\hbar$, the optical conductivity is treated within the Mattis-Bardeen theory \cite{Mattis_PhysREv1958}. A plot with theoretical $\sigma_1(\omega,T)$ is shown in Fig.~\ref{pic:stripline}(b). In the superconducting state the energy gap forms around the Fermi surface, and the states within the energy gap are transferred to the edge of the energy gap, forming Van Hove singularities. A depiction of the density of states in a fully gapped superconductor is shown in the inset of Fig.~\ref{pic:sigma_main}(e). For frequencies below the energy gap, only the thermally excited quasiparticles can absorb energy and contribute to $\sigma_1$. This contribution to $\sigma_1$ we denote as $\sigma_1^\mathrm{th}$. At very low frequencies, the temperature dependence $\sigma_1(T)$ exhibits a so-called coherence peak, a broad maximum at temperatures slightly below $T_\mathrm{c}$, which reflects the Van Hove singularities in the density of states. $\sigma_1(T)$ with a coherence peak is shown as green line in Fig~\ref{pic:stripline}(b). If the frequency is greater than $2\Delta(T)$, quasiparticles can be excited across the energy gap, leading to an extra absorption channel $\sigma_1^\mathrm{ph}$. The total conductivity $\sigma_1=\sigma_1^\mathrm{th}+\sigma_1^\mathrm{ph}$ then exhibits upturns as a function of $\omega$ or $T$ when the excitation frequency matches the energy gap $2\Delta(T)$. The energy gap can be seen as a sharp kink in the $\sigma_1(\omega,T)$ manifold and is marked as an orange line in Fig.~\ref{pic:stripline}(b). 
By projecting this kink down to the $\omega-T$-plane (red line in Fig.~\ref{pic:stripline}(b)), the temperature dependence and magnitude of the energy gap can be observed directly. 
Fig.~\ref{pic:sigma_main}(a) and (b) show the frequency dependence of $\sigma_1$ and $\sigma_2$, respectively, for temperatures above and below $T_\mathrm{c}$. Here the conductivity at each frequency was normalized to the normal-state conductivity $\sigma_n$ measured above $T_\mathrm{c}$.
When entering the superconducting state $T<T_\mathrm{c}=0.47$~K, we observe a reduction of $\sigma_1(\omega)$ for the highest five frequencies, whereas for the lowest, at $f_0=3.35$~GHz, $\sigma_1$ increases first when lowering the temperature below $T_\mathrm{c}$. 
This behavior is explained by the energy gap opening, and spectral weight of $\sigma_1(\omega)$ is shifted to lower frequencies, resulting in a reduced $\sigma_1(\omega)$ around the energy gap.
In Fig~\ref{pic:sigma_main}(b) the frequency dependence of $\sigma_2$ for various temperatures is shown. $\sigma_1$ and $\sigma_2$ are connected by the Kramers-Kronig-relations. 
At very low temperatures and at frequencies below the energy gap, the overall behavior of $\sigma_1$ is dominated by the $\delta(\omega)$-peak at zero frequency caused by the superfluid condensate whereas quasiparticle contributions to $\sigma_1$ vanish. 
As Kramers-Kronig transform of the $\delta$-peak, a $1/f$-frequency dependence is expected for $\sigma_2(\omega)$ in the superconducting state, which is indicated by the dashed line in Fig.~\ref{pic:sigma_main}(b) and experimentally observed for frequencies below the gap.

Fig.~\ref{pic:sigma_main}(c),(d) show the temperature dependence of the optical conductivity at different frequencies. At the lowest shown frequency, $\sigma_1(T)$ exhibits a pronounced upturn just below $T_\mathrm{c}$ as the sample enters the superconducting state. This is the before mentioned coherence peak and reflects the enhanced density of states at the edges of the energy gap. 
Although BCS theory and the Mattis-Bardeen formalism were developed in the late 1950's, the coherence peak in the optical conductivity was observed only in the 1990's and remains in the focus of microwave experiments on superconductors \cite{Holczer_SolStat1991, Marsiglio_PhysRevB1994,Klein_PhysRevB1994, Jin2003, Steinberg_PRB2008}. 
As we go up in frequency, $\sigma_1(T)$ in the superconducting state decreases compared to lower frequencies due to the reduced number of states the thermal quasiparticles can be excited into. At the highest measured frequency of 38.91~GHz, $\sigma_1(T)$ does not vanish at low temperatures, since the excitation frequency is above the zero-temperature energy gap $2\Delta_0$, and therefore breaking of quasiparticles is possible even for lowest temperatures.

Panel (e) in Fig.~\ref{pic:sigma_main} shows the imaginary part $\sigma_2(T)$ of the optical conductivity, which is mainly related to the superfluid in the superconducting state at low frequencies. At low temperatures $\sigma_2(T)$ becomes constant, because then all quasiparticles are condensed into the superfluid.

The qualitative behavior of $\sigma(T)$ fits quite well with the behavior predicted by Mattis-Bardeen theory, which is shown in the panels (e) and (f) of Fig~\ref{pic:sigma_main} as comparison to panels (d) and (e).

\subsection{Superconducting energy gap}

\begin{figure}
	\includegraphics[width=0.9\linewidth]{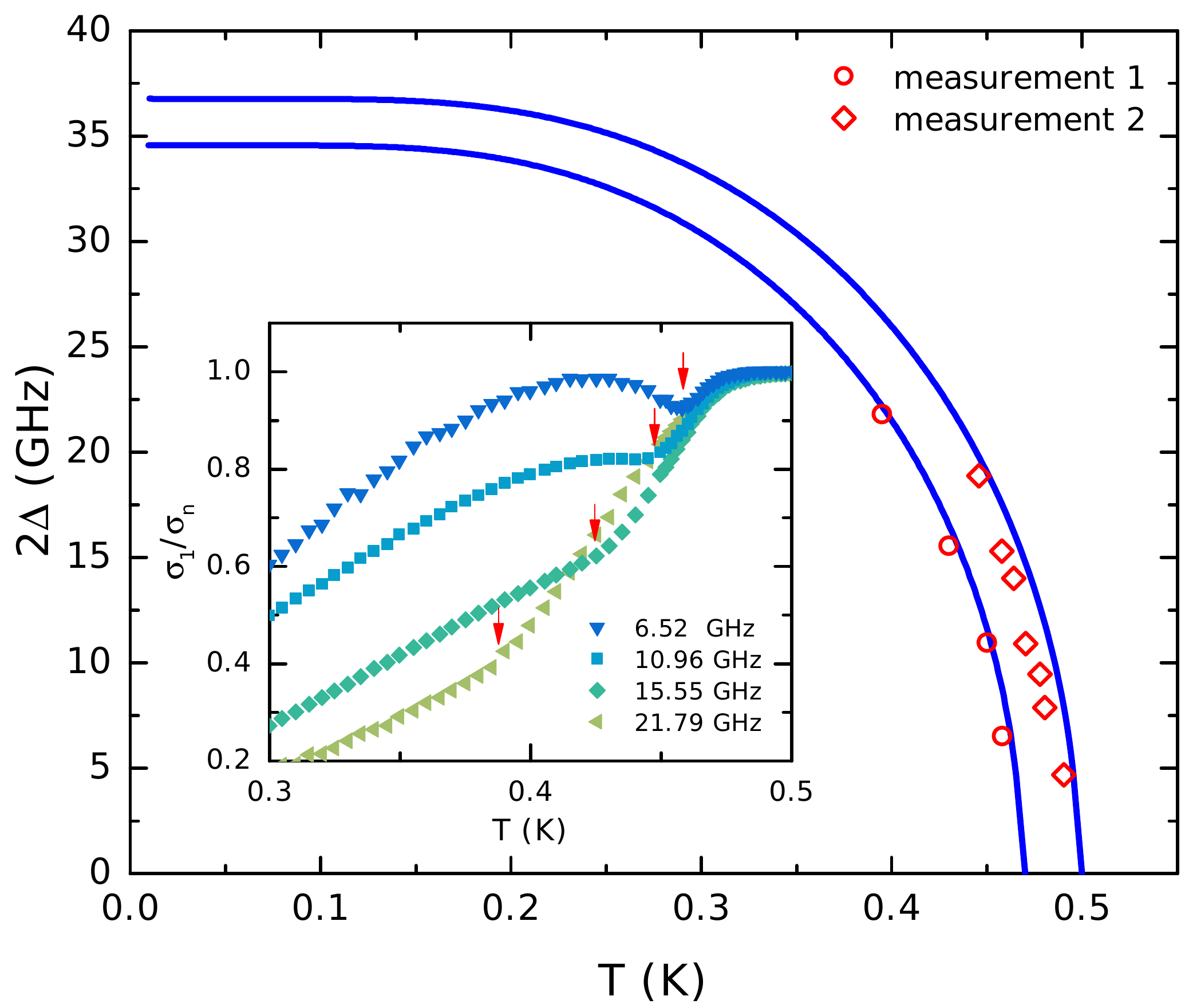}
	\caption{ (Inset) As a function of temperature $\sigma_1$ shows distinct kinks (marked by arrows) indicating the temperature where $\Delta(T)$ matches the applied microwave frequency. (Main) Circles and diamonds indicate the energy gap for the two measurements as determined by the procedure of the inset. The lines are BCS predictions with $ 2\Delta/k_BT=3.53 $.}
	\label{pic:DeltavsT}
\end{figure}
In principle one can quantify the superconducting energy gap $2\Delta$ from measured data by fitting the $\sigma(\omega)$ spectra to the theoretical expectation \cite{Pronin_PRB1998, Xi2010, Pracht_PRB2012}, in the simplest case based on the Mattis-Bardeen formalism \cite{Pracht2013, Mattis_PhysREv1958, Zimmermann1991}.
The accuracy of determining the energy gap this way depends on the experimental frequency resolution, which in our case is not sufficient for meaningful fits with $\Delta$ as free parameter. We therefore follow a different approach by evaluating $\sigma_1(T)$ at a fixed frequency. Here we expect an abrupt change in the temperature dependence once the excitation frequency matches the energy gap $2\Delta$, as visible in Fig.~\ref{pic:stripline}(b). Since $2\Delta$ is temperature dependent, different excitation energies will match the energy gap at different temperatures. The advantage of looking at $\sigma_1(T)$ is that our temperature resolution is much higher than our frequency resolution.

For the intermediate frequencies ($f=6.52$~GHz - $21.79$~GHz) we can easily observe this abrupt change in temperature dependence, as marked by the red arrows in the inset of Fig.~\ref{pic:DeltavsT}, which reproduces data of Fig.~\ref{pic:sigma_main}(c) close to $T_\mathrm{c}$. The main panel of Fig.~\ref{pic:DeltavsT} shows the values for the energy gap $2\Delta(T)$ determined by this method for both measurements. The blue lines are the temperature dependence of the energy gap predicted by weak coupling BCS theory with  $2\Delta_0/k_\mathrm{B}T_\mathrm{c}\approx 3.53$, which properly describes our data for both measurements.
We thus obtain values of 71$\mu$eV and 76$\mu$eV for $\Delta_0$ for the two measurements.

\subsection{Superfluid density, penetration depth, and scattering rate}

\begin{figure}
	\includegraphics[width=0.9\linewidth]{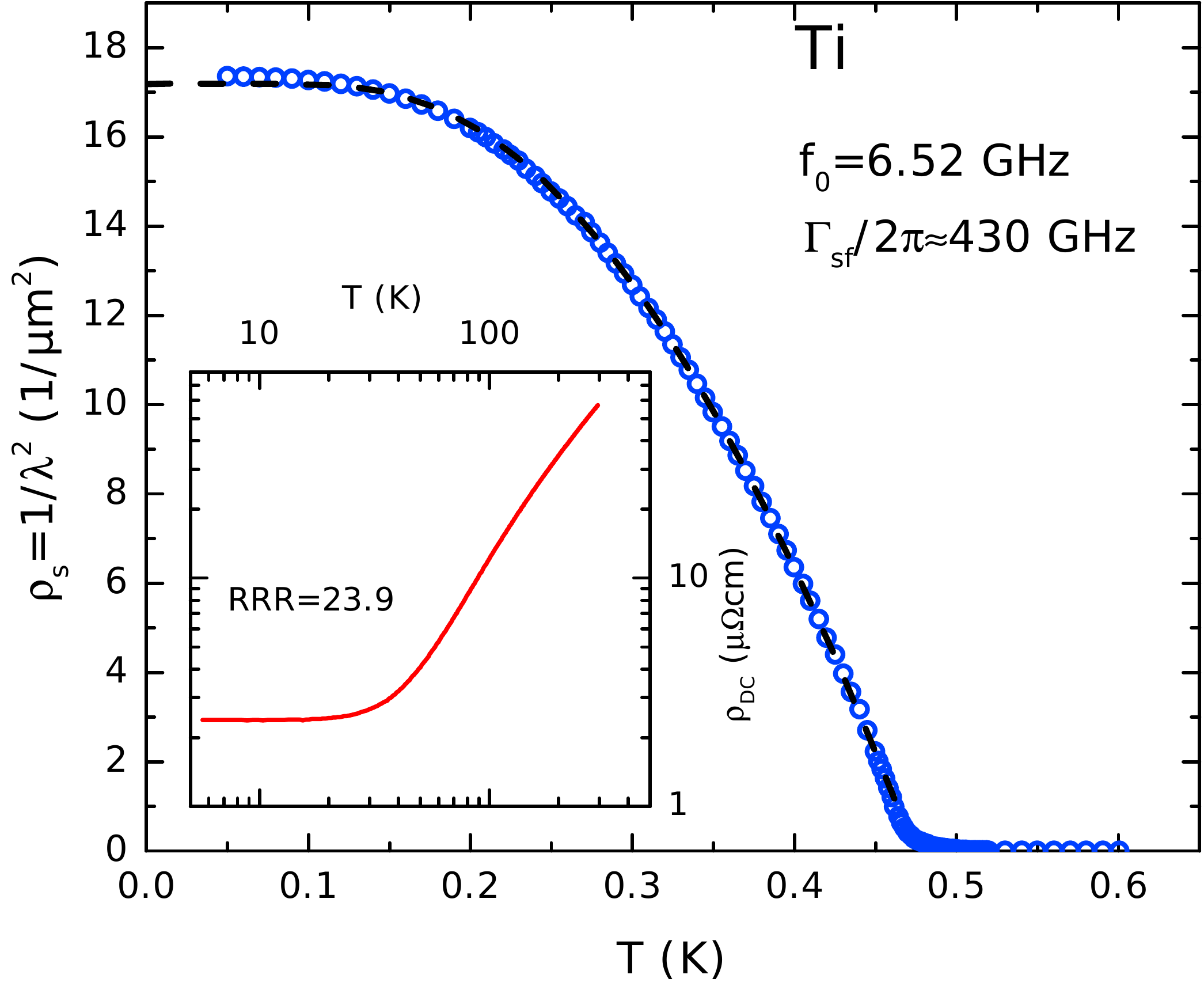}
	\caption{ Temperature dependence of the superfluid density $\rho_s$ calculated from the $\sigma_2$ data obtained at $f_0=6.52$~GHz. The black dashed line is a fit for BCS-weak-coupling  superfluid density in the presence of disorder, with $T_\mathrm{c}$ and the scattering rate $\Gamma/2\pi$ as fit parameters. (Inset) Temperature dependence of the DC resistivity, measured in four-point geometry.
The flattening of $\rho_\mathrm{DC}$ at about 30~K and the low RRR ratio of 23.9 indicate strong scattering present in the sample.}
	\label{pic:sup_density}
\end{figure}

So far we have only considered the response of the thermal quasiparticles and the breaking of Cooper pairs, which contribute to $\sigma_1$. The response of the superfluid is encoded in the out-of-phase response $\sigma_2$. At low frequencies the superfluid density $\rho_s$ is connected to $\sigma_2$ via \cite{Tinkham_book}
\begin{equation}
\rho_s(T)=1/\lambda(T)^2=\lim_{\omega\rightarrow 0}\mu_0 \omega \sigma_2(\omega,T)
\label{eq:supdensity}
\end{equation}
In the clean case where $\Gamma/2\pi\rightarrow 0$, the spectral weight in $\sigma_1(\omega)$ available to condense into the superfluid is the full spectral weight of the normal-state Drude peak, which is given in terms of the plasma frequency by $\rho_\mathrm{s00}=\mu_0\epsilon_0\omega_p^2$. 
With increasing scattering, spectral weight is shifted to higher frequencies, out of the range where it condenses into the superfluid. 
Therefore an increase of scattering causes a decrease of the superfluid density. In the presence of scattering, the temperature dependence of the superfluid density can be calculated by
\begin{equation}
\rho_s(T)= \rho_\mathrm{s00} 2\pi k_{B} T\sum_{\omega_n>0}^{\infty}\frac{1}{\sqrt{\omega_n^2 + \Delta(T)^2} + \frac{\Gamma_\mathrm{sf}}{2\hbar}}\frac{\Delta(T)^2}{\omega_n^2 + \Delta(T)^2}
\label{eq:sup_dens_dirty}
\end{equation}
where $\omega_n=2\pi k_B T(n+1/2)$ are the Matsubara frequencies \cite{Nam_PhysRev1967}. The dashed black line in the main panel of Fig.~\ref{pic:sup_density} is a fit of Eq.~\ref{eq:sup_dens_dirty} to the measured superfluid density obtained via Eq.~\ref{eq:supdensity} from our $\sigma_2$ data at 6.52~GHz, where $T_\mathrm{c}$ (which enters Eq.~\ref{eq:sup_dens_dirty} via the temperature dependence of $\Delta$, which is assumed BCS-like) and $\Gamma_\mathrm{sf}$ were fit parameters. From the fit we determine $T_\mathrm{c}=0.474$~K and $\Gamma_\mathrm{sf}/2\pi=430$~GHz. Comparing $\Gamma_\mathrm{sf}/2\pi$ with the scattering rate from resistivity $\Gamma_\rho/2\pi=490$~GHz, we find them in good agreement. The fit allows us to extract the zero-temperature superfluid density $\rho_s(0\mathrm{~K})$ and consequently the zero temperature penetration depth $\lambda_0=1/\sqrt{\rho_s (0\mathrm{~K})}=241~\mathrm{nm}$.

Next we would like to comment on the change in $T_\mathrm{c}$ after polishing, and the absence of multigap features in our data. (All theory descriptions above consider only a single superconducting gap, and all our data, most notably the temperature-dependent penetration depth in Fig.~\ref{pic:sup_density}, are fully consistent with this assumption.) According to Anderson's theorem, scattering leads to an averaging of the energy gap over the Fermi surface \cite{Anderson_JPhysChemSolids1959}. As revealed by our measurement of the superfluid density, scattering plays a substantial role for the superconductivity in Ti. Furthermore, in Ref.~\cite{Peruzzi_NucPhysB1999} it is shown that impurities can change the transition temperature of Ti by a factor of 2. Recent resistivity measurements indicate an anisotropy of the Debye frequency, which would lead to an anisotropy of the energy gap in the case of phononic coupling \cite{Chandra_PhysicaB2017}. Depending on the amount of defect scattering (which we may have modified by polishing, since we only probe within a few hundred nm from the surface), the maximum energy gap on the Fermi surface has different values, and therefore the samples will vary in $T_\mathrm{c}$. Furthermore, as mentioned in the introduction, Ti exhibits several electronic bands crossing the Fermi energy, making Ti potentially a multiband and multigap superconductor. Depending on the strength of interband scattering, the potentially different energy gaps of different bands will average, leaving a single superconducting energy gap throughout the complete Fermi surface \cite{Thiemann_PRL2018}. That we do not detect any signs of multiple energy gaps present in superconducting Ti therefore is consistent with the observed scattering rate that is much larger than the energy gap.

\subsection{Behavior in magnetic field}
\begin{figure}
\includegraphics[width=\linewidth]{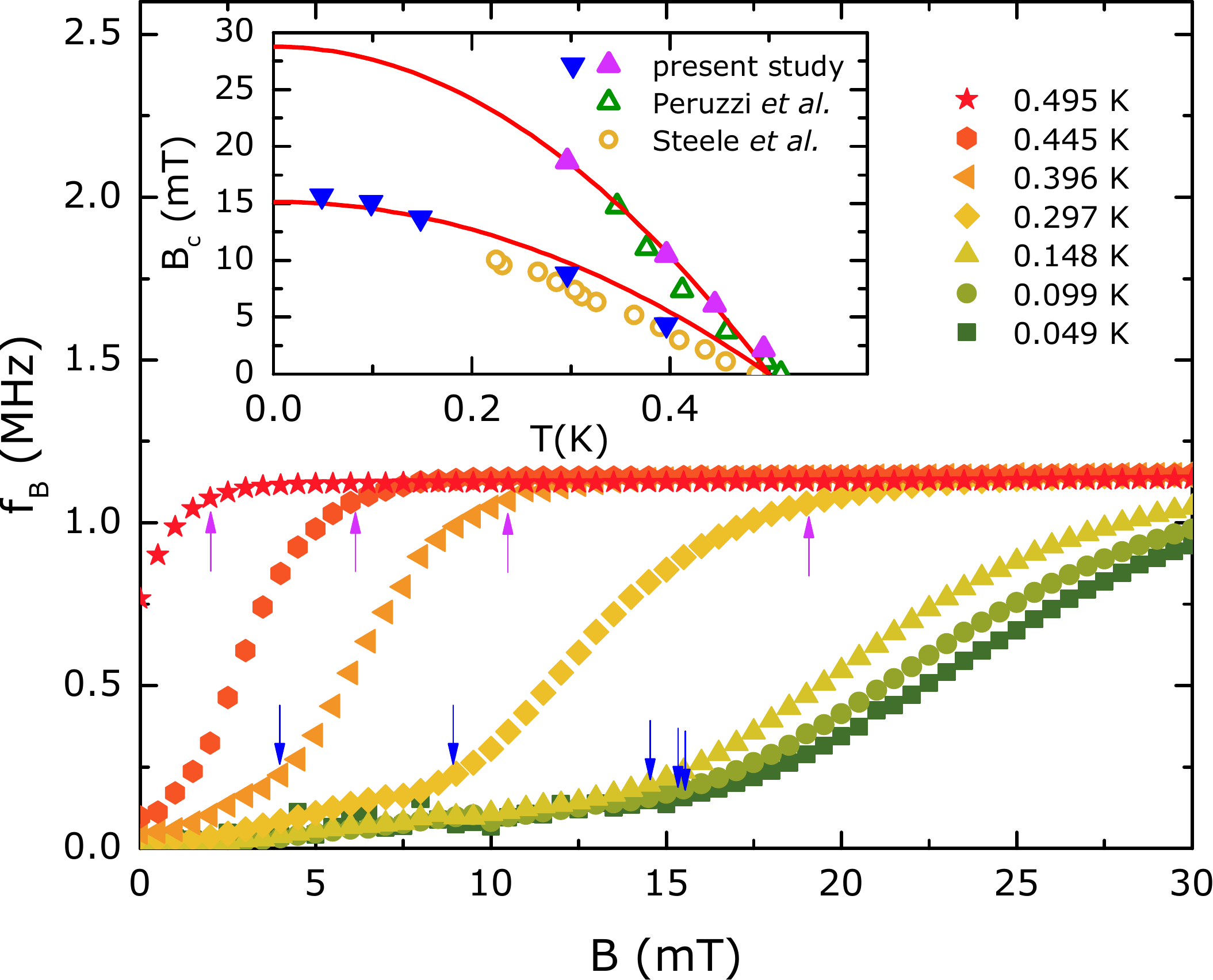}
\caption{Magnetic field dependence of the resonator bandwidth $f_B$ measured at different temperatures up to $T_\mathrm{c}$. The blue and purple arrows indicate the beginning and end of the superconducting transition. The resulting temperature dependence of the two field scales is plotted in the inset, together with the data reported by Steele \textit{et al.} and Peruzzi \textit{et al.} \cite{Steele_PhysRev1953, Peruzzi_Metr2000}. The red lines are fits according to Eq.~\ref{eq:BcvsT}.}
\label{pic:BcvsT} 
\end{figure}

Titanium is a type I superconductor, i.e.\ any magnetic field is expelled from the interior of the sample until the external applied field reaches the value $B_\mathrm{c}$. Fig.~\ref{pic:BcvsT} shows the magnetic field dependence of the measured resonator bandwidth $f_B$ for different temperatures up to $T_\mathrm{c}$. Interestingly the superconducting transition driven by magnetic field appears much broader than the one in zero field driven by temperature. Similarly, a broadening of the superconducting transition in Ti with increasing static external magnetic field has been reported in Ref.~\cite{Peruzzi_NucPhysB1999}. To quantify this effect, we read out the start and end of the superconducting transition, marked by blue and purple arrows respectively in Fig.~\ref{pic:BcvsT}. The resulting temperature dependence of these characteristic magnetic fields is shown in the inset of Fig.~\ref{pic:BcvsT} as blue and purple triangles. In previous studies on the temperature dependence of the critical field of superconducting Ti with a similar $T_\mathrm{c}$, rather diverse values for the critical fields have been reported \cite{Peruzzi_NucPhysB1999,Steele_PhysRev1953}. Comparing our two field scales with those in literature we find good correspondence with both, which suggests that the broad transition of superconducting Ti is responsible for the large scattering reported in literature. The solid lines in Fig.~\ref{pic:BcvsT} are fits to \cite{Tinkham_book}
\begin{equation}
B_c(T)=B_c(0)\left(1-\left(\frac{T}{T_\mathrm{c}}\right)^2\right).
\label{eq:BcvsT}
\end{equation}
The fits properly describe our experimental data and thus confirm the conventional parabolic temperature dependence that one expects for the critical magnetic field. Unfortunately, we cannot explicitly assign either of the two field scales to a the conventional definition of $B_c$: microwave spectroscopy on type I superconductors previously found a similar effect for Pb that was ascribed to surface superconductivity, but that observation was in a rather narrow field range compared to the present observation (and even narrower or absent for Sn) \cite{Ebensperger2016}, and thus it is not clear to which extent this explanation can also be applied to the present case of Ti.

\section{summary}
We performed resonant microwave measurements on superconducting titanium using stripline resonators and determined the optical conductivity $\sigma(\omega,T)$ in frequency and temperature ranges 3-40~GHz and 40-600~mK.  
Qualitatively the frequency and temperature dependence of $\sigma(\omega,T)$ is in excellent agreement with the predictions of the Mattis-Bardeen theory. We can observe unique signatures of the energy gap $2\Delta$ in the frequency and temperature dependence of $\sigma_1$, and from the latter we can determine the temperature dependence of $2\Delta$, which nicely follows the BCS temperature dependence in the weak coupling limit with a ratio of $2\Delta/k_BT_\mathrm{c}\approx 3.53$. Therefore we conclude that Ti is a BCS-like superconductor.

The temperature dependence of the superfluid density, obtained from the imaginary part of the optical conductivity, allows us to determine the absolute value of the scattering rate $\Gamma_\mathrm{sf}/2\pi=430$~GHz, which compares well with the value determined from resistivity $\Gamma_\rho/2\pi= 490$~GHz. These scattering rates clearly indicate Ti being a superconductor in the dirty limit. From the absolute values of the superfluid density we determine the zero-temperature penetration depth $\lambda_0=241$~nm.

\section*{Acknowledgments}
We thank G. Untereiner for resonator and sample preparation, A. L\"ohle for support with the DC measurement, and D. M. Broun and N. R. Lee-Hone for fruitful discussions.  M. T. thankfully acknowledges financial support by the Carl-Zeiss-Stiftung. We thankfully acknowledge financial support by the Deutsche Forschungsgemeinschaft (DFG).

\end{document}